\documentstyle[preprint,pra,aps]{revtex}
\begin{document}
\author{Dmitriy Kilin and Michael Schreiber}
\address{
  Institut f\"ur Physik, 
  Technische Universit\"at, D-09107 Chemnitz, Germany }
\title{Decoherence for phase-sensitive relaxation}
\date{\today}
\maketitle
\begin{abstract} 
  It is shown that the vibrational wave packet relaxation of initially
  coherent (displaced) states as well as the quantum superposition of
  coherent states in heat bathes with the different spectral densities 
  exhibit a number of peculiarities 
  compared with the cases of linear phase-sensitive 
  relaxation, quadratic relaxation,
  and relaxational dynamics described with the use of Zurek's
  environmentally-induced pointer basis. 
 A strong dependence of the relaxation rate 
 on the position of the spectral density
 maximum of the bath is found. 
  The differences can be used
  for the discrimination of the mechanisms of the molecule-environment
  interactions.

\noindent
PACS number(s): 82.29.Rp, 33.80.Be, 32.80.Bx, 31.70.Hq

\end{abstract}


\section{Introduction}
The behavior of many quantum systems strongly depends on their
interaction with the environment. It is important to take this
interaction into account to realistically describe systems like, e.g.,
vibrational levels in a big molecule, the quantized mode of an
electromagnetic field, or a trapped ion.  The rapid development of
experimental techniques in these and other branches of physics and
chemistry leads to an increased interest in theoretical descriptions
of possible experiments by numerical calculations.  Additionally our
calculations, which concern such systems, allow us to regard the still
existing question about the border between classical and quantum
effects from a new point of view. For the systems, which are described
here, the border to disappearance of quantum effects has been
estimated.

One of the fundamental questions of quantum physics is to understand
why the general principle of superposition works very well in
microscopic physics but leads to paradox situations in macroscopic
physics such as the Schr\"{o}dinger cat paradox \cite{1} where the cat
can exist in a superposition of the states dead and alive. One
possible explanation of the paradox and the nonobserving of a
macroscopic superposition is that \cite{2} systems are never
completely isolated but interact with an environment, that contains a
large number of degrees of freedom. Interactions with the environment
lead to continuous loss of coherence and drive the system from a 
superposition into a statistical classical mixture.

The interest in the decoherence problem
is explained not only by its relation to the
fundamental question: ''Where is the borderline between the
macroscopic world of classical physics and microscopic phenomena ruled
by quantum mechanics?'', but also by the increasing significance
of potential practical applications of quantum mechanics,
such as quantum computation and cryptography [3, 4].

There are a number of propositions how to create the
superposition states in mesoscopic systems, or systems that have both
macroscopic and microscopic features. Representative examples
are the superposition of two coherent states of an harmonic oscillator
\begin{equation}
  \left| \alpha ,\phi \right\rangle =N^{-1}\left( \left| \alpha
    \right\rangle +e^{i\phi }\left| -\alpha \right\rangle \right)
  \label{1}
\end{equation}
for a relatively large amplitude ($\alpha \sim 3\div 5$). Here,
$\left| \alpha \right\rangle $ is a coherent state and 
$N=\left[
          2+2\cos \phi \exp \left( -2
                                    \left| 
                                           \alpha 
                                    \right|^2
                              \right) 
   \right]^{1/2}$
is a normalization constant. These states have been observed
recently for the intracavity microwave field \cite{5} and for motional
states of a trapped ion \cite{6}. Additionally, it has been predicted
that superpositions of coherent states of molecular vibrations could
be prepared by appropriately exciting a molecule with two short laser
pulses \cite{7} and the practical possibilities of realizing such an
experiment have been discussed \cite{8}.  In this scheme the quantum
interference would survive on a picosecond time scale, which is
characteristic for molecular vibrations.  

From the theoretical point
of view, quantum decoherence has been studied extensively
[2, 9-16].
Most efforts focused on the
decoherence of the harmonic oscillator states due to the coupling to a
heat bath, consisting of a large number of oscillators representing
the environment. The system is usually described on the basis of the
master equation for a reduced density operator. There are two general
approaches for this method. One adopts the Markov approximation
together with the rotating wave approximation (RWA). That means that
all details of the complex system-environment interactions are
neglected and relaxation is described by characteristic decay
constants. The system-bath interaction is phase-insensitive because of
the RWA. In another approach, according to Zurek \cite{2}, the
coupling with the environment singles out a preferred set of states,
called ''the pointer basis''. Only vectors of this basis survive the
quantum dynamics.  The vectors of the pointer basis are the
eigenvectors of operators, which commute with the (full) interaction
Hamiltonian of the system. This basis depends on the form of the
coupling. Very often this pointer basis consists of the eigenstates of
the coordinate operator. The density operator describing the system
evolves to diagonal form in the pointer basis, which is usually
connected to the disappearance of quantum interference. The two
approaches give different pictures of the same decoherence processes.

One of the goals of this contribution is to present a consistent
analysis of the decoherence on the basis of a density matrix approach
starting from von Neumann's equation for the density matrix of 
the whole system,
i.e. the microscopic quantum system and the ''macroscopic''
environment.
\section{Generalized master equation}
Let us consider a single molecule vibrating in a one-dimensional
harmonic potential. The molecule interacts with a number of harmonic
oscillators modeling the environment. In the interaction
Hamiltonian
\begin{equation}
  H_{\mathrm{ES}}=\hbar \sum_\xi K_\xi \left( b_\xi ^{+}+b_\xi \right)
  \left( a+a^{+}\right)  \label{2}
\end{equation}
$a$ ($a^{+}$) are annihilation (creation) operators of molecular
vibrations with the frequency $\omega $, $b_\xi$ ($b_\xi ^{+}$) operators
for the environmental vibrations having the frequencies $\omega _\xi
$. $K_\xi $ is the coupling between them.  Starting from the von
Neumann equation
\begin{equation}
  \dot{\rho}=-\frac i\hbar \left[ H,\rho \right] , \label{3}
\end{equation}
where the  Hamiltonian
\begin{equation}
  H=H_{\mathrm{S}}+H_{\mathrm{E}}+H_{\mathrm{ES}} \label{4}
\end{equation}
contains the molecular system 
$H_{\mathrm{S}}=\hbar \omega \left(a^{+}a+1/2\right) $ 
and the environment
$H_{\mathrm{E}}=\sum_\xi \hbar \omega _\xi 
                      \left( 
                            b_\xi ^{+}b_\xi +1/2
                      \right) $,
one can formally rewrite the equation
for the reduced density matrix
$\sigma =\mathrm{tr}_{\mathrm{E}} \left( \rho \right) $, 
which is averaged over the environmental states [17, 18]
\begin{equation}
  \dot{\sigma}=-\frac i\hbar \left[ H_{\mathrm{S}},\sigma \right]
 + e^{-{i H_{\mathrm{S}}t}/{\hbar}}\dot{D}(t,0)
   \left(D{(t,0)}\right)^{-1} 
   e^{ {i H_{\mathrm{S}}t}/{\hbar}}\sigma ,
  \label{5}
\end{equation}
where
\begin{equation}
  D{(t,0)}={\rm tr}\left( \hat{T}\exp \int_0^td\tau \left( -\frac
      i\hbar L(\tau )\right) \cdot \rho _E(0)\right) \label{6}
\end{equation}
is an evolution operator, averaged over the initial states of the
environment.
\begin{equation}
  L(\tau )=\left[ H_{\mathrm{ES}}(\tau ),\qquad \right] \label{7}
\end{equation}
is the Liouville operator in the interaction representation and
$\hat{T}$ is the operator of chronological ordering. Supposing that
the initial states of the bath oscillators are thermalized  
($\rho_\xi (0) \sim \exp (
                          -\hbar \omega _\xi b_\xi ^{+}b_\xi
                           /k_{\mathrm{B}}T
                          )$),
and restricting a cumulant expansion of $D$ to
the second order cumulant [17, 18],
we obtain the non-Markovian
master equation
\begin{equation}
  \dot{\sigma}=-i\omega \left[ a^{+}a,\sigma \right] +R\sigma ,
  \label{8}
\end{equation}
where the action of the relaxation operator $R$ is defined by
\begin{eqnarray}
  R\sigma 
       &=&  \left[ 
                  \left( 
                     A_{n+1}+A_n^{+}
                  \right) 
               \sigma ,
               a^{+}+a 
            \right]
          + \left[ 
                a^{+}+a,
                \sigma 
                   \left( 
                      A_n+A_{n+1}^{+}
                   \right) 
            \right].
  \label{8a}
\end{eqnarray}
Here, the operators $A_n$ and $A_{n+1}$ are defined by linear
combinations of the operators $a$ and $a^{+}$ as
\begin{eqnarray}
     A_n &=&       \gamma _n  \left( t\right) a
           +\tilde{\gamma}_n  \left(t\right) a^{+},
                                          \label{10} \\ 
 A_{n+1} &=&       \gamma _{n+1} \left( t\right) a
           +\tilde{\gamma}_{n+1} \left( t\right)a^{+}, 
\label{10}
\end{eqnarray}
with the functions
\begin{equation}
        \gamma _{n+1}(t)
            =\int_0^tR_{n+1}(\tau )d\tau 
            =\sum_\xi K_\xi^2 (n_\xi+1)
                  \frac
                       {e^{-i(\omega _\xi -\omega)t}-1}
                       {-i(\omega _\xi -\omega )},
\label{10a}
\end{equation}
\begin{equation}
        \gamma _n(t)
            =\int_0^tR_n(\tau )d\tau 
            =\sum_\xi K_\xi ^2 n_\xi
                  \frac
                       {e^{-i(\omega _\xi -\omega )t}-1}
                       {-i(\omega _\xi -\omega )},
\label{10b}
\end{equation}
\begin{equation}
\tilde{\gamma} _{n+1}(t)
           =\int_0^tR_{n+1}(\tau ) e^{-2i\omega\tau }d\tau 
           =\sum_\xi K_\xi ^2 (n_\xi+1)
                 \frac
                      {e^{-i(\omega _\xi +\omega )t}-1}
                      {-i(\omega _\xi +\omega)}, 
\label{10c}
\end{equation}
\begin{equation}
\tilde{\gamma} _n(t)
           =\int_0^tR_n(\tau ) e^{-2i\omega \tau}d\tau
           =\sum_\xi K_\xi ^2 n_\xi
                 \frac
                      {e^{-i(\omega _\xi +\omega )t}-1}
                      {-i(\omega _\xi +\omega )}. 
\label{10d}
\end{equation}
Note, that $R_n$ and $R_{n+1}$ are the correlation functions of the
environmental perturbations 
$R_n(t)=\sum_\xi K_\xi ^2 n_\xi
                 \exp 
                    \left( 
                         -i\left( 
                                \omega _\xi -\omega 
                           \right) 
                           \tau 
                    \right) $,
where $n_\xi$ denotes the number of quanta in the bath mode 
with the frequency $\omega_\xi$.

The functions $\gamma$ correspond to the friction coefficient in the classical limit. 
The first two coefficients
$\gamma_n$ and $\gamma_{n+1}$
strongly depend on the coupling constant $K_\xi$
for frequencies $\omega _\xi \simeq \omega $ and 
on the number of quanta in the bath modes 
with the same frequencies, 
whilst the coefficients 
$\tilde{\gamma}_n$ and $\tilde{\gamma}_{n+1}$
are very small for all frequencies.

The obtained master equation (\ref{8})
describes different stages of vibrational relaxation. 
The initial stage is defined by a period of time smaller than the
correlation time $\tau _c$ of the environmental perturbations. This
time can roughly be estimated from the width $\Delta \omega $ of the
perturbation spectrum $K_\xi ^2$, for $\tau _c\sim 1/\Delta \omega $.
For such small times one can write
the master equation in the following form
\begin{equation}
  \dot{\sigma} 
        = -i\omega \left[ a^{+}a,\sigma \right] \label{11}  
            +\Gamma
                \left\{
                    \left[
                        \left( a^{+}+a \right) \sigma ,a^{+}+a 
                    \right]
                   +\left[ 
                         a^{+}+a ,\sigma\left( a^{+}+a \right) 
                    \right]  
                \right\} ,
\end{equation}
where 
$\Gamma 
=\sum_\xi K_\xi ^2\left( 2 n_\xi+1\right) $ 
is a real constant. As follows from Eq.~(\ref{11}),
the pointer basis for this step of relaxation is defined by the
eigenstates of the position operator 
$\hat{Q}\sim a^{+}+a $. Another period of time, for which the form of the
relaxation operator $R$ according to Eq.~(\ref{8a}) is universal, is the
kinetic stage, where $t\gg \tau _c$ and the Markov approximation
becomes applicable. In this stage the master equation has the form
\begin{eqnarray}
 \dot{\sigma} &=&-i\omega \left[ a^{+}a,\sigma \right] \label{12} \\ 
  && + \gamma 
         \left\{ 
             \left[ 
                \left( 
                   (n+1) a+na^{+}
                \right) \sigma, 
                a^{+}+a 
             \right]   
           + \left[ 
                a^{+}+a,
                \sigma
                \left(
                   (n+1) a+na^{+}
                \right) 
              \right]
         \right\}, \nonumber
\end{eqnarray}
where $ \gamma =\pi K^2 g $ is the decay rate of the vibrational
amplitude.
Here $n=n_\xi$, $K=K_\xi$ and the density of bath states
$g=g(\omega_\xi)$ are evaluated at the frequency $\omega=\omega_\xi$
of the selected oscillator.
It should be stressed that Eq.~(\ref{12}) differs from the usual
master equation for a damped harmonic oscillator for derivation of
which the RWA is applied [5-8, 10-13]. This phase-sensitive
relaxation leads to new effects: classical squeezing and a
decrease of the effective harmonic oscillator frequency \cite{18}.
In between there is a time interval, where relaxation
is specific and depends on the particular spectrum of $K_\xi ^2$.
\section{ Analytical solution for wave packet dynamics}
The solution of the equation
of motion of the reduced density matrix can be conveniently
found using the characteristic function formalism, introduced in \cite{19},
which enables us
to use the differential operators
 $\frac{\partial}{\partial \lambda} $ and 
 $\frac{\partial}{\partial \lambda^*} $
instead of $ a^+ $ and $ a $.
 Multiplying 
 both sides
of Eq.(\ref{8}) with  the factor 
$f=\exp{(\lambda a^+ )} \exp{(-\lambda^* a)} $
and taking the trace 
one can rearrange all terms into such a form that 
$a$ and $ a^+ $ precede the appropriate exponent.
For this operation we change the order of operators
using the expression
$ a \exp{(\lambda a^+)} = \exp{(\lambda a^+)} \left( a+\lambda  \right) $
to make the transformation
$ a^+ \exp{(\lambda a^+)}  = 
                    \frac{\partial}{\partial \lambda} \exp{(\lambda a^+)}$.
After that every term can be represented by the
normally ordered characteristic function
$ F={\rm tr}(f \sigma)$
upon which one of the differential operators acts.
We obtain, e.g.,
${\rm tr} \left( 
            \left[ 
                  a^+ a, \sigma 
            \right] f 
          \right) 
   = \left( 
        \lambda^* \frac{\partial}{\partial \lambda^*}
      - \lambda   \frac{\partial}{\partial \lambda}
     \right) F  $,
${\rm tr} \left( 
            \left[ 
                  a^+ \sigma, a 
            \right] f 
          \right) 
   = \lambda^*
     \left( 
        \frac{\partial}{\partial \lambda}
      - \lambda^*
     \right) F  $, and
${\rm tr} \left( 
            \left[ 
                  a \sigma, a^+ 
            \right] f 
          \right) 
   =  - \lambda^* \frac{\partial}{\partial \lambda^*} F  $.
Such manipulations lead us finally to the  
${\it complex}$%
-valued partial differential equation:
\begin{eqnarray}
  \dot{F} 
      &=& -  \left[ 
                  i\omega \lambda ^{*}
                  + \mu
                  \left( 
                     \lambda +\lambda ^{*}
                  \right) 
             \right] 
             \frac{\partial}{\partial \lambda^{*}}
             F \nonumber
         +   \left[ 
                 i\omega \lambda 
                 -\mu^{*}
                  \left( 
                    \lambda+\lambda ^{*}
                  \right) 
              \right] 
              \frac{\partial}{\partial \lambda} F \nonumber \\ 
&&       -    \left( \lambda +\lambda ^{*}\right) 
              \left( 
                  \nu \lambda^{*}+\nu^{*}\lambda 
              \right) F, \label{14}
\end{eqnarray}
where
\begin{eqnarray}
  \mu (t) &=&       \gamma_{n+1}(t)
            +\tilde{\gamma}_n^*(t)-\nu^{*}(t), \nonumber \\
  \nu (t) &=&       \gamma_n^{*}(t)
            +\tilde{\gamma}_{n+1}(t) \label{15}
\end{eqnarray}
are relaxation functions. We can solve Eq.~(\ref{14}) by using the
integral representation for the characteristic function 
$F\left( \lambda,\lambda ^{*},0\right) 
  ={\rm tr}\left( 
                  e^{ a^{+} \lambda    }
                  e^{-a     \lambda^{*}}
              \right) $ 
which formally allows us  to describe the
nondiagonal density matrix. Below the notation 
$\int \int d\alpha d\beta c(\alpha ,\beta )
          =\left\langle 
                \alpha \right. \left| \beta
           \right\rangle $ 
is adopted. An initial characteristic function
\begin{equation}
  F\left( \lambda ,\lambda ^{*},0\right) 
   =\int \int d\alpha d\beta
              e^{\alpha \lambda -\beta \lambda ^{*}}
              c(\alpha ,\beta ) 
\label{16}
\end{equation}
will evolve in accordance with 
Eq.(\ref{14}) as
\begin{equation}
  F\left( \lambda ,\lambda ^{*},t\right) 
    =\int \int d\alpha d\beta
         \exp 
            \left(
               \sum_{m,n}
                   K_{mn}^{(\alpha ,\beta )}(t)
                   \lambda^m\left( -\lambda ^{*}\right) ^n
            \right) 
            c(\alpha ,\beta ).
  \label{17}
\end{equation}
We restrict the cumulant expansion to the second order, i.e.
$ m+n \le 2 $
.
For a wide class of initial states
(coherent, thermal, squeezed, etc.) high order cumulants
vanish and our approximation becomes exact.
The cumulants
could hold  nondiagonal information,
such as the density matrix, in relevant cases we stress it
with the upper index $ (\alpha) $ or $ (\beta) $. 

The functions $%
K_{mn}(t)$ in Eq.(\ref{17}) are given by the
solutions of the sets of equations
\begin{eqnarray}
      \dot{K}_{01}^{(\beta )} 
             &=& -  \left( 
                        i\omega +  \mu 
                     \right) K_{01}^{(\beta)}
                                                     +  \mu^{*}
                                                       K_{10}^{(\alpha)},        
\nonumber \\ 
      \dot{K}_{10}^{(\alpha )} 
             &=&     \left( 
                         i\omega - \mu ^{*}
                     \right) K_{10}^{(\alpha)}
                                                    +  \mu
                                                      K_{01}^{(\beta)},
\label{18}
\end{eqnarray}
\begin{eqnarray}
  \dot{K}_{11} 
        &=&  2{\rm  Re}\nu
             -2 \left( {\rm  Re}\mu \right) 
                        K_{11}
             +2 \mu     K_{02}
             +2 \mu^{*} K_{20}
, \nonumber \\
  \dot{K}_{20} 
        &=& -  \nu^{*}+\mu K_{11}
            +2 \left( 
                  i\omega - \mu^{*}
               \right) K_{20}, \label{19} \\
  \dot{K}_{02} 
        &=&  - \nu 
              + \mu^{*} K_{11}
             - 2 \left( 
                   i\omega - \mu 
                \right) K_{02}, \nonumber
\end{eqnarray}
with the initial values
\begin{equation}
\begin{array}{c}
  {K}_{10}^{(\alpha )}(t=0) = \alpha , \\ 
  {K}_{01}^{(\beta )} (t=0) = \beta  , \\ 
  K_{11}\left( t=0\right) =K_{20}\left( t=0\right) =K_{02}\left(
    t=0\right) =0.
\end{array}
\label{20}
\end{equation}
For the special case $\beta =\alpha ^{*}$, these initial conditions
represent a coherent state with amplitude $\alpha $. This solution
can be used for the construction of wave packets in different
representations.  Here, we will discuss the coordinate representation,
in particular
the dependence of the probability density $P$ on the vibrational
coordinate $Q$ and on time
\begin{equation}
  P\left( Q,t\right) =\frac 1{2\pi }\int_{-\infty }^\infty d\lambda
  e^{-i\lambda Q}\chi \left( \lambda ,t\right) , \label{21}
\end{equation}
where
\begin{equation}
  \chi \left( \lambda ,t\right) 
     =   {\rm tr}\left( e^{i\lambda \left(
          a^{+}+a\right) }\sigma (t)\right) =e^{-\lambda^2/2}
          F\left( i\lambda ,-i\lambda ,t\right) \label{22}
\end{equation}
is a characteristic function for the position operator. Evaluating the
integral (\ref{21}) we finally obtain
\begin{equation}
  P\left( Q,t\right) =\int \int d\alpha d\beta P^{(\alpha ,\beta
    )}\left( Q,t\right) c(\alpha ,\beta ), \label{23}
\end{equation}
where
\begin{equation}
P^{(\alpha ,\beta )}\left( Q,t\right) =\frac 1{2\sqrt{\pi V\left( t\right) }%
}\exp \left\{ 
          -\frac
              {
               \left[ 
                    Q-Q^{(\alpha ,\beta )}\left( t\right) 
               \right]^2
              }
              {
               4V\left( t\right) 
              }
      \right\} \label{24}
\end{equation}
and
\begin{equation}
\begin{array}{c}
  Q^{(\alpha ,\beta )}\left( t\right) =K_{10}^{(\alpha )}\left(
    t\right) +K_{01}^{(\beta )}\left( t\right) , \\ V\left( t\right)
  =\frac 12+K_{11}\left( t\right) +K_{20}\left( t\right) +K_{02}\left(
    t\right) .
\end{array}
\label{25}
\end{equation}
For the case $\beta =\alpha ^{*}$ the function 
$Q^{\left( \alpha,\alpha ^{*}\right) }$ 
denotes the expectation values of the
coordinate operator of the coherent state, $V$ 
is the broadening of the Gaussian packet (\ref{24}). 
The distribution $P$ can be used for the
investigation of relaxation dynamics for any initial state of
molecular vibration, but it is best suited for studying the
evolution of states prepared as a superposition of coherent states.
Below, we will discuss the relaxation dynamics of initially 
coherent states and the superposition (\ref{1}) of two coherent states.
\section{Relaxation dynamics of coherent states and their superpositions}
\subsection{Coherent states}

Because of the importance of this case we will use it to investigate
the relaxation dynamics described by the master equation (\ref{8}).
For coherently exited states $\left| \alpha _0\right\rangle $ the
initial characteristic function is 
$F\left( \lambda ,\lambda^{*},0\right) 
  =\exp 
     \left(
         \alpha _0^{*}\lambda - \alpha _0\lambda^{*}
     \right) $. 
Thus, the initial values for 
$K_{10}^{(\alpha)}\left( t\right) $ 
and $K_{01}^{(\beta )}\left( t\right) $ 
are $\alpha _0^{*}$ and $\alpha _0$.  
In the first stage of relaxation, when $t\ll \tau _c$, 
the relaxation functions are  
$
\mu\left( t\right)=0$,  
$    \nu \left( t\right) 
    =\Gamma
t $.
Therefore, the solution of the system of Eqs. (\ref{18}, \ref{19}) 
gives
\begin{equation}
\begin{array}{c}
  Q^{(\alpha _0^{*},\alpha _0)}\left( t\right) 
    =2{\rm Re}\left(
                  \alpha_0 e^{-i\omega t}
              \right) \\ 
  V\left( t\right) 
    =\frac12 + \Gamma
t^2.
\end{array}
\label{26}
\end{equation}
Even in this early stage there is a small quadratic broadening of the
wave packet $P^{\left( \alpha _0^{*},\alpha _0\right) }\left(
  Q,t\right) $ without changing its mean amplitude.  After the
intermediate stage of relaxation the solution of the system goes into
the Markovian stage of relaxation, where the master equation (\ref{12}) works.
For this stage the solution of Eqs. (\ref{18}, \ref{19}) reads
\begin{eqnarray}
  Q^{\left( \alpha _0^{*},\alpha _0\right) }\left( t\right)
  &=& 2{\rm Re}\left( \alpha _0z\left( t\right) \right) e^{-\gamma t},
\nonumber \\
V\left( t\right) &=& \frac 12+n-ne^{-2\gamma t}\left[ 1+\left( \frac \gamma {%
      \tilde{\omega}}\right) ^2\left( 1-\cos 2\tilde{\omega}t\right)
  +\frac \gamma {\tilde{\omega}}\sin 2\tilde{\omega}t\right] ,
\label{28}
\end{eqnarray}
where
\begin{eqnarray}
z\left( t\right) & = & \cos \tilde{\omega}t+(\gamma /\tilde{\omega})\sin \tilde{%
  \omega}t+i(\omega /\tilde{\omega})\sin \tilde{\omega}t,
\nonumber
\\
\tilde{\omega}   & = & \sqrt{\omega ^2-\gamma ^2}, \label{29}
\\
               n & = & \left[\exp (\hbar \omega/kT)-1\right]^{-1}.
\nonumber
\end{eqnarray}



\noindent
Equation~(\ref{29}) demonstrates the decrease of the effective harmonic
oscillator frequency due to the phase-dependent interaction with the
bath. 
The broadening of the wave packet $V(t)$ can be also considered
as uncertainty of the coordinate. We would like to discuss both its increase
and oscillation, described by Eq.(31) and displayed in Fig. 1.
The increase of $V(t)$ appears due to absorbtion of quanta from the 
heat bath. Such an effect is rather clear and can be obtained 
already using RWA. The oscillation of the broadening 
deserves more intent attention. Some authors [21, 22]
have predicted the oscillation of second moments, 
using Green functions formalism. One feature of prediction [22] 
is that it assumes initially a squeezed state.
Our prediction, represented in explicit form in Eq.~(31) goes further.
We would like to underline that in our case such an effect
appears even for the usual coherent state as the initial one.
In spite of the fact that this oscillation does not 
present quantum squeezing, 
because the width is never smaller than the ground state width,
we stress the derivation
of wave packet broadening oscillation in Eq.~(31) 
as one important achievement of the present paper.
\subsection{Superposition of two coherent states}
For the initial superposition of two coherent states (\ref{1}) the
normally ordered characteristic function consists of four terms:
\begin{equation}
  F\left( \lambda ,\lambda ^{*},0\right) 
   =N^{-2}
         \left[ 
                F_{\alpha^{*},\alpha }
            +   F_{-\alpha ^{*},-\alpha }
            +   e^{-2\left| \alpha \right| ^2}
                     \left( 
                          e^{i\phi }F_{\alpha ^{*},-\alpha }
                       +  e^{-i\phi}F_{-\alpha ^{*},\alpha }
                     \right) 
         \right]
\label{30}
\end{equation}
with $F_{\alpha ,\beta }=e^{ \alpha \lambda - \beta \lambda ^{*} }$. 
The first terms describe the mixture of two coherent states, 
$\left|    \alpha \right\rangle \left\langle   \alpha \right| 
+ \left|  - \alpha\right\rangle \left\langle  - \alpha \right| $. 
The last two terms correspond to the quantum interference, 
i.e., they reflect the coherent properties of the superposition.  
In accordance with Eqs.(\ref{18})--(\ref{25}) 
and the equation of motion  (\ref{11}) of the reduced
density matrix this initial state evolves as
\begin{equation}
P\left( Q,t\right) 
  =  \frac 1{N^2}
          \left[
              P^{\left( \alpha ^{*},\alpha \right) }\left( Q,t\right) 
          +   P^{\left(-\alpha ^{*},-\alpha \right) }\left( Q,t\right) 
          \right] 
     + P_{int} \left( Q,t \right)
\end{equation}
with the interference term
\begin{equation}
 P_{int} \left( Q,t \right)
  =  \frac 1{N^2}
              e^{-2\left|\alpha \right| ^2}
                   \left[ 
                         e^{i\phi }
                         P^{\left(\alpha ^{*},-\alpha \right) }
                          \left( Q,t\right)
                      +  e^{-i\phi }
                         P^{\left( -\alpha^{*},\alpha \right) }
                          \left( Q,t\right)  
                   \right] 
\label{31} 
\end{equation}
where 
$P^{\left( \alpha ,\beta \right) }\left( Q,t\right) $ 
is given by Eq.(\ref{24}) with
\begin{equation}
         Q^{\left( \alpha ^{*},\alpha \right) }\left( t\right) 
   =  -  Q^{\left(-\alpha ^{*},-\alpha \right) }\left( t\right) 
   =     2 {\rm  Re}
                \left(\alpha z(t) \right) 
                e^{-\gamma t}, 
\label{32}
\end{equation}
\begin{equation}
         Q^{\left( \alpha ^{*},-\alpha \right) }\left( t\right) 
   =  -  Q^{\left(
      -  \alpha ^{*},\alpha \right) }\left( t\right) 
   =     2i {\rm  Im}
                 \left(\alpha z\left( t\right) \right) 
                 e^{-\gamma t}. 
\label{33}
\end{equation}
$V\left( t\right) $ and $z\left( t\right) $ are defined by
Eqs.(\ref{28}, \ref{29}). Rewriting the real part of
Eq.(\ref{31}) we finally obtain
\begin{eqnarray}
  P_{int}\left( Q,t\right) 
   &=&\frac 
        1
        {N^2}
        \frac 
                   1
                   {\sqrt{\pi V\left( t\right) }}
              \exp \left(
                      - 2\left| \alpha \right| ^2
                      + \frac
                             {  4 \left[ 
                                      {\rm  Im}
                                            \left( \alpha z(t)\right) 
                                  \right]^2
                                  e^{-2\gamma t}
                               -  Q^2}
                             {4 V(t) }
                   \right) \label{34} \\ 
& &\times \Biggl. \cos 
                      \left( 
                            \phi 
                          - Q 
                            \frac
                                 {{\rm  Im} ( \alpha z(t)) }
                                 {V(t)}
                             e^{-\gamma t}
                      \right)
         . \nonumber
\end{eqnarray}\noindent
\noindent 
Figure 2 illustrates, how the superpositional state ~(1)
evolves in time in accordance with Eq. ~(34). 
It follows from Eq.~(38), that the interference term describing
quantum coherence in the system is only significant when
\begin{equation}
\frac{\left[ {\rm  Im}\left( \alpha z\left( t\right) \right) \right] ^2}{%
  V\left( t\right) }e^{-2\gamma t}\approx 2\left| \alpha \right| ^2.
\label{35}
\end{equation}
This is true for $\gamma t\ll 1$, and moreover, when the two wave packets
 of the state (\ref{1}) come close together
(i.e., at the moment when $ z\left( t_i\right) \approx
i$). Expanding Eq.(\ref{35}) into 
a series at these points and taking
into account $\gamma t_i\ll 1$, it is easy to see that 
\begin{equation}
P_{int}
 \sim
  \exp \left( 
          -   2 \left| \alpha \right| ^2
          +   \frac
                  {\left[ {\rm  Im}
                               \left(
                                  \alpha z\left( t_i\right) 
                               \right) 
                   \right] ^2}
                  {V \left( t_i \right) }
              \left( 1-2\gamma t_i \right) 
      \right). 
\label{36}
\end{equation}
The decoherence is due to two reasons:
The spreading 
$V\left( t_i\right) =1/2+\Delta V\left(t_i\right) n$
of the wave packet 
due to thermal excitations by the bath,
and amplitude decoherence. The second means, 
that even for the case 
$ n=0 $ and 
$  
      {\rm  Im}
             \left(
                \alpha z\left( t_i\right) 
             \right) 
  =  \left| \alpha \right|$ 
quantum interference disappears exponentially with the rate
\begin{equation}
  t_{\mathrm{dec}}^{-1}\simeq 2\left| \alpha \right| ^2\gamma
  \label{37}
\end{equation}
This result obtained by Zurek \cite{2} is the main reason why quantum
interference is difficult to observe in the mesoscopic and macroscopic
world. For example, a physical system with mass 1g 
in a superposition state with a separation of 1cm 
shows a ratio of relaxation and decoherence time scales of $10^{40}$.  
Even if our measuring device is able to reflect the quantum 
properties of the microsystem, nevertheless
objectification \cite{20} occurs due to the coupling 
between the meter and the environment.
The fundamental result of Eq.(\ref{37}) is obtained no matter which
approach is used, e.g. the RWA or Zurek's pointer basis approach or
the self-consistent description of the present paper. However, the
time dependence of the superposition terms of the distribution of
Eq.(38) differs a little bit, which can be seen in Fig. 3.  As
quantum interference is more sensitive, we have used it for
comparison of three different approaches to the present problem. Figure
3 shows the difference between the time evolutions, which result from Eq.
(\ref{26}), from Eq.(\ref{28}), and from the corresponding result of
the RWA approach.

These differences arise due to the fact that during the relaxation
there is no constant pointer basis for all steps of the
evolution. As follows from Eqs. (\ref{11}, \ref{12}), this basis
changes from the position eigenstate basis in Eq. (\ref{11}) to
the more complicated basis in Eq. (\ref{12}).
\subsection{Partial conservation of superposition}
As mentioned above the interference term (35) 
 of the wave packet decays inevitably.
However, there exist proposals, how 
to slow down this process,
reducing the consequence of the interaction.
In particular, one can consider a situation, 
when the majority of the bath modes 
have an off-resonance frequency in 
respect to the selected system.
In such a system the processes which 
are induced by the 
interaction with bath, namely loss of 
amplitude and phase must be delayed.
The hope to observe the conservation 
of a superpositional state
as a consequence of the mentioned properties
is the reason to put attention on such a system.
As a simplest example we take the quantum system
surrounded by
harmonic oscillators with 
twice the system frequency $\omega_\xi=2\omega$.
It provides processes, when the system 
loses 2 quanta and the bath obtains one
quantum. 
Although this is still a resonace situation,
we demonstrate below that it leads to
unusual behavior compared to the usual
situation $\omega_\xi=\omega$, discussed above.
Describing such a situation 
in the RWA
we rewrite the interaction Hamiltonian  (2) 
in the following form
\begin{equation}
H_I=\hbar \sum_{\xi} K_\xi 
\left(
b^+_{\xi}a^2 +b_\xi (a^+)^2
\right).
\end{equation}
Applying then the above mentioned formalism of the 
evolution operator (6) we obtain again the equation  (8)
for the reduced density matrix $\sigma$
of the selected system,
but with some changes of the relaxational part (9)
\begin{eqnarray}
R \sigma &=& \Gamma (n_\xi + 1)
         \left\{ 
             \left[  
                    a^2
                      \sigma, 
                (a^+)^2
             \right]   
           + \left[ 
                (a^+)^2,
                \sigma
                    a^2 
              \right]
         \right\} \nonumber \\
 &+& \Gamma n_\xi
         \left\{ 
              \left[ 
                   (a^+)^2 \sigma, a^2 
             \right]   
           + \left[ 
                a^2, \sigma (a^+)^2 
              \right]
         \right\} ,
\end{eqnarray}
where $ \Gamma =\pi K_\xi^2 g_\xi $ is the decay rate of the vibrational
amplitude.
Here, the number of quanta in the bath mode
$n_\xi$, 
the coupling function
$K_\xi$, 
and the density of bath states
$g_\xi$ are evaluated at 
$\omega_\xi=2 \omega$, i.e., at
the double frequency 
of the selected oscillator.

Rewritten in the basis of eigenstates  
$\left| n \right\rangle$
of the unperturbed oscillator
this master equation
contains only linear combinations of such terms as
$\sigma_{m,n}=\left\langle m \right|\sigma \left|n\right\rangle$,  
$\sigma_{m+2,n+2}$, and   $\sigma_{m-2,n-2}$.
It ensures, in effect, a possibility to distinguish
even and odd initial states of the system.
The odd excited state $\left| 1 \right\rangle$ 
cannot relax to the ground state $\left| 0 \right\rangle$,
but the even excited state $\left| 2 \right\rangle$ can. 

The evolution of the system 
after preparation under
different initial conditions
was simulated numerically.
The equations of motion of the  density matrix  elements
are integrated using a fourth order 
Runge-Kutta algorithm with stepsize control.
To make the set of differential equations a finite one
we restrict the number 
of levels by $m,n \leq 20$.

Investigating the evolution of the initially 
coherent state of such a system
we have concentrated our attention on some detail,
which was not seen before.
One of the representative examples is the behavior 
of the initially coherent state of the system.
For comparison with the usual behavior we show
the time dependence of the mean value of the coordinate
 in Fig. 4a.
The mean value in 
the usual system decreases with a constant rate.
The same initial value of the system coupled to a 
bath with $\omega_\xi=2 \omega$ shows a fast decrement 
in the first stage and almost no decrement
afterwards.

The question about the influence of the type 
of the bath on the evolution of the superpositional state
is of special importance. At a temperature 
of $k_B T = 2 \hbar \omega / {\rm ln}3$,
corresponding to $n(2\omega)=0.5$,
we have simulated the evolution
of the same superpositional states twice,
for $\omega_\xi=\omega$ and $\omega_\xi=2\omega$.
The coordinate representation 
of the wave packets is presented in Fig. 4b and 4c.
The same value of the relaxation $\gamma=\Gamma$ yields, 
however,  different results.
In the system coupled to the usual bath 
the quantum interference disappears
already during the first period,
while the  amplitude decreases only slightly.
In the other case, the system coupled to
the bath with $\omega_\xi=2\omega$ leaves the interference 
almost unchanged, although
a fast reduction of the amplitude occurs.

Therefore, the second system partially conserves 
the quantum superpositional state.
The experimental investigation of systems displaying
such properties [2] is necessary,
both for extracting typical parameter ranges for theoretical models
 and for practical applications like
quantum computation and quantum cryptography.

The described example leads 
us to the conclusion that 
the second type of environment is more preferable.

\section{Conclusions}
Starting from von Neumann's equation for a vibrational oscillator
interacting with the environment modeled by a set of independent
harmonic oscillators we derived a non-Markovian master equation, which
has been solved analytically. 
For two types of the bath with maximum of the spectral density 
near the system frequency and near twice the system frequency and
for two different initial states, namely
a coherent state and a superposition of coherent states, the wave
packet dynamics in coordinate representation have been analyzed. It has 
been shown that wave packet dynamics demonstrate 
either
''classical squeezing''
and a decrease of the effective vibrational oscillator frequency due
to the phase-dependent interaction with the bath
or a time-dependent relaxation rate, distinct for even and odd states,
and partial conservation of quantum superposition
due to the quadratic interaction with the bath.  
The decoherence
also shows differences compared to the usual damping processes
adopting RWA and to the description using the pointer basis for
decoherence processes. We conclude that there is no permanent pointer
basis for the decay. There are two universal stages of relaxation: the
coherence time scale stage and the Markovian stage of relaxation, both
having different pointer bases. We believe that the proposed
method can be applied for other initial states and different
couplings with the environment in real existing quantum systems, which is
important in the light of recent achievements in single molecule
spectroscopy, trapped ion states engineering, and quantum computation.

\newpage
\begin{figure}[ht]\centering
\caption[fig1]
{ Dynamics of the wave packet $P\left( Q,t\right) $ (left plot) and
  its variance $V$(right plot) for $\gamma =0.1\omega$, $
  k_{\mathrm{B}}T=3\hbar \omega$, $Q_0=4$.
\label{1}
}
\end{figure}
\begin{figure}[ht]\centering
\caption[fig2]
{ Time evolution of the superposition of coherent states for
  $\omega =1$, $\gamma =0.01\omega$, $n=0$, $\alpha =2$, $\phi =\frac
  \pi 2$.
\label{2}}
\end{figure}
\begin{figure}[ht]
\caption[fig3]
{ Time dependence of  $P_{int} \left( Q,t\right)$ 
  for $\omega=1$,  $\gamma=0.25\omega$, $n=0.4$,
  $\alpha=2$, $\phi=0$, $Q_0=0$.
  The solid line represents the case ~(31), 
  the boxes represent RWA, the bullets represent 
  earliest-time analysis ~(30).}
\label{3}
\end{figure}

\begin{figure}[p]\centering
  \parbox{7.3cm}
  {\rule{0cm}{0cm}
\caption[fig2]
{ Influence of different baths.
  (a) Evolution of the coherent state:
  Mean value of the coordinate,
  for $Q_0=-2.2$.
  Diamonds: system coupled to
  the bath with $\omega_\xi=\omega$, $\gamma=0.15\omega$.
  Solid line: system coupled to
  the bath with $\omega_\xi=2\omega$, $\Gamma=0.5\omega$. 
  (b) Time evolution of the superposition of coherent states 
  with initial separation $2Q_0=8$ 
  for $\omega_\xi=\omega =1$, $\gamma= 0.005\omega$, 
  $n=1.36$.
  (c) Same as (b), but for $\omega_\xi=2 \omega$ with
  $\Gamma=0.005\omega$, $n_\xi=0.5$.
\label{2}
}}
\end{figure}
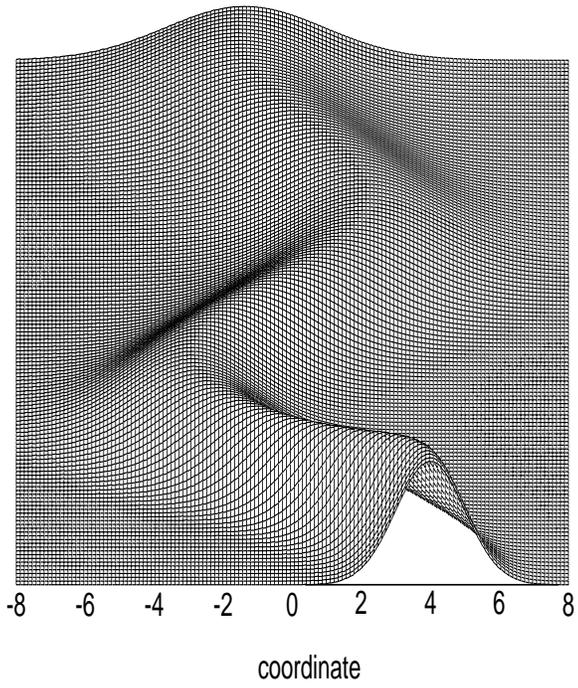
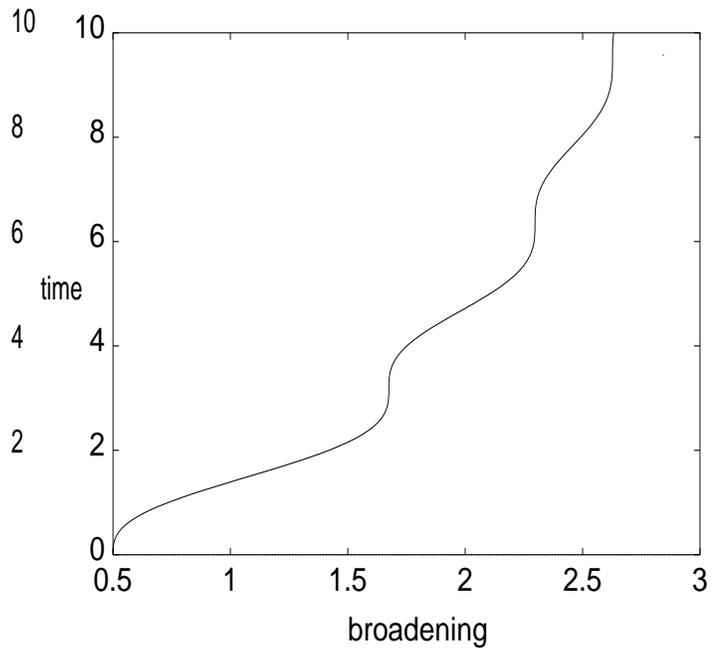
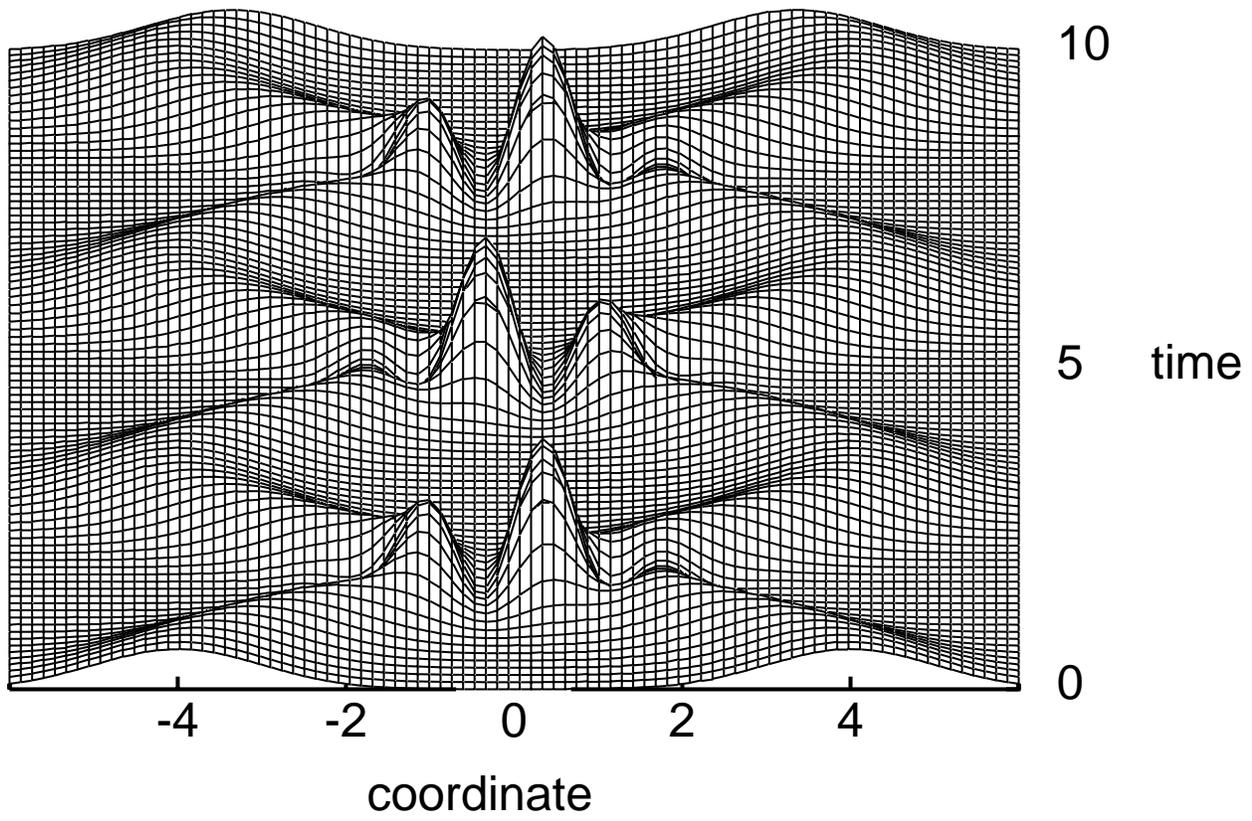
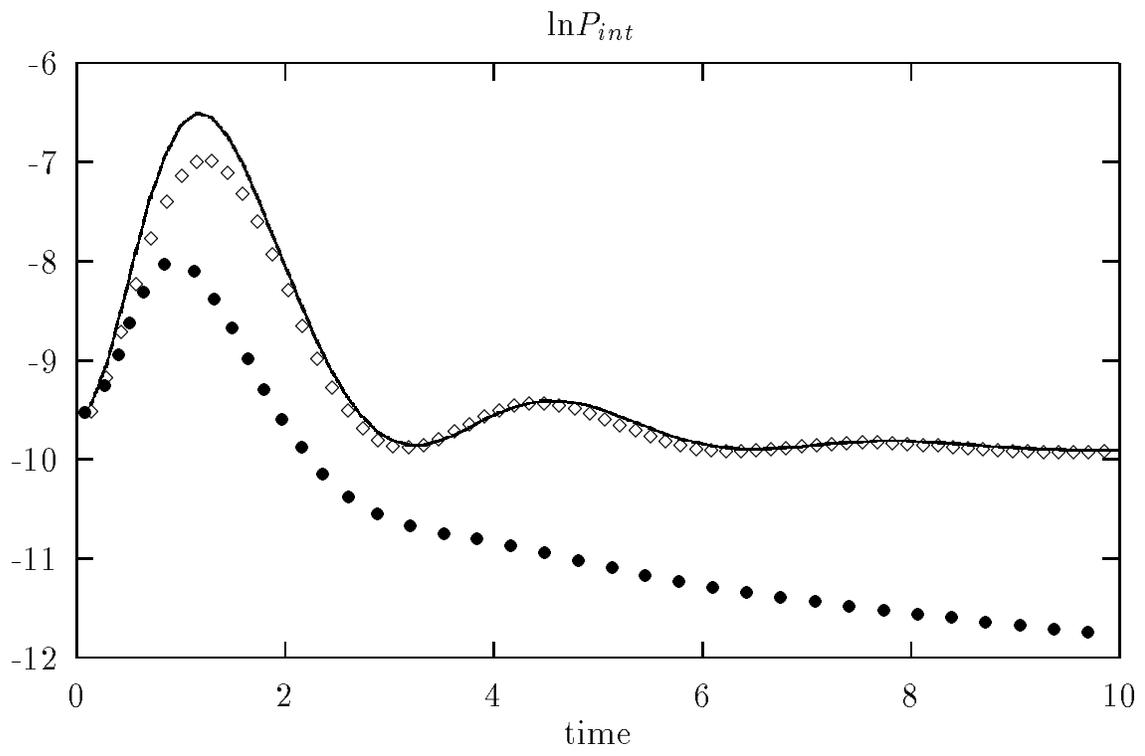
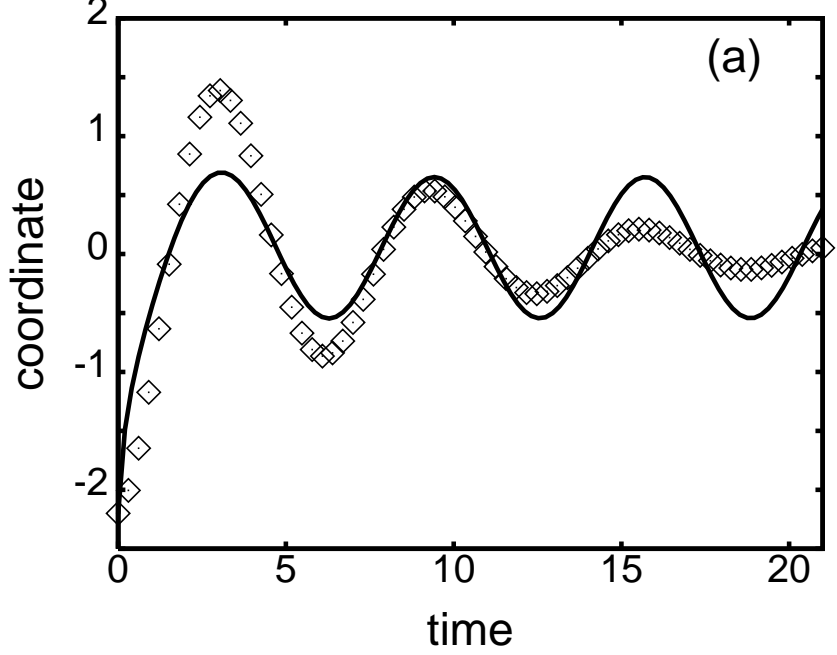
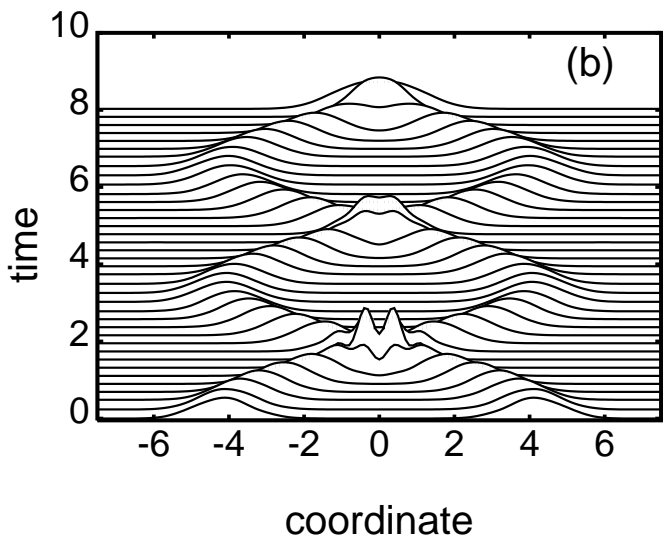
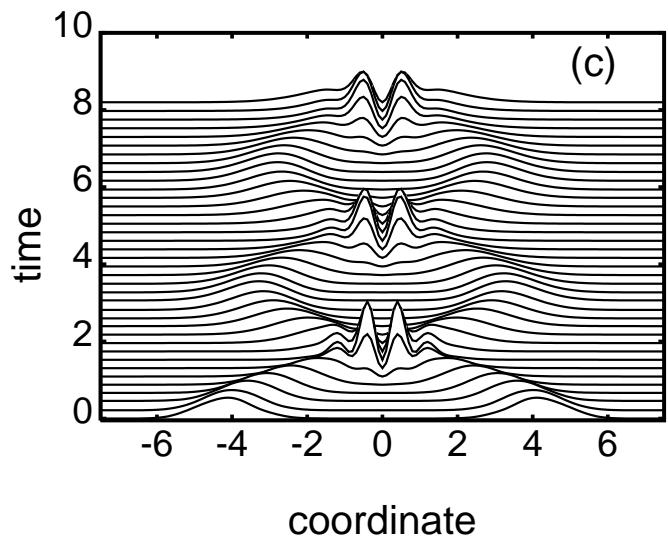
\end{document}